\documentstyle[sprocl,epsf]{article}

 \catcode`@=11
\def\@cite#1#2{\unskip\nobreak\relax
    \def\@tempa{$\m@th^{\hbox{\the\scriptfont0 #1}}$}%
    \futurelet\@tempc\@citexx}
\def\@citexx{\ifx.\@tempc\let\@tempd=\@citepunct\else
    \ifx,\@tempc\let\@tempd=\@citepunct\else
    \let\@tempd=\@tempa\fi\fi\@tempd}
\def\@citepunct{\@tempc\edef\@sf{\spacefactor=\the\spacefactor\relax}\@tempa
    \@sf\@gobble}
 
\def\citenum#1{{\def\@cite##1##2{##1}\cite{#1}}}
\def\citea#1{\@cite{#1}{}}
\catcode`@=12

\begin{document}

\font\fortssbx=cmssbx10 scaled \magstep1
\hbox to \hsize{
\hbox{\fortssbx University of Wisconsin - Madison}
\hfill$\vcenter{\hbox{\bf MADPH-97-1004}
                \hbox{July 1997}}$ }

\bigskip

\title{SEMI-LEPTONIC \boldmath{$B$} DECAY\footnote{Talk presented at the {\it Conference on $B$ Physics and CP Violation}, March 24--27, 1997, Honolulu, Hawaii.}}
\author{\vskip-3ex M. G. OLSSON}
\address{Physics Department, University of Wisconsin, Madison, WI 53706}
                                
\maketitle
\abstracts{
The definitions and properties of elastic and inelastic
               Isgur-Wise form factors are reviewed.  The nature of the
               ``missing" exclusive semi-leptonic decay fraction is
                identified. The role that completeness sum rules
                play in relating form factors, spectroscopy, and model                building is outlined.}

Semi-leptonic $B$ decay occupies a unique position in hadron physics.  On one hand it is the most numerous decay channel and on the other it provides                                           an almost ideal application of the Heavy Quark Symmetry (HQS).  As such it stands alone as a unique opportunity to make detailed tests of HQS\cite{isgurwise} and Heavy Quark Effective Theory (HQET)\cite{reviews}.  Within this framework universal Isgur-Wise (IW) form factors require careful model building (or perhaps lattice simulation) to make complete predictions.

To place the gross features of semi-leptonic decay in perspective we quote the Particle Data Group\cite{pdg} branching ratio results for the inclusive and the two largest exclusive $B^0$ semi-leptonic decays,
\begin{equation}
\begin{array}{lcr}
{\rm Br}[B\to X \ell\nu] &=& 10.3\pm1.0 \% \,,\\
{\rm Br}[B\to D^*\ell\nu] &=& 4.6\pm0.3 \% \,,\\
{\rm Br}[B\to D\ell\nu] &=& 1.9\pm0.5 \%\,.
\end{array}
\label{Br}
\end{equation}
By subtraction one finds that the semi-leptonic decay not into a single $D$ or $D^*$ is
\begin{equation}
{\rm Br}[B\to (\mbox{not $D$ or $D^*$})\ell\nu]=3.8\pm1.3 \% \,.
\label{notsingleD}
\end{equation}
One of the themes of this talk will be to address the significance of this number.

\section{Heavy Quark Symmetry and the definitions of Isgur-Wise functions}

In the HQS limit the heavy quark becomes a static source of color electric field and its color magnetic moment vanishes.  A heavy-light meson can thus be labeled by its Light Degrees of Freedom (LDF) angular momentum $j$ and the meson parity $P$.  There are thus two degenerate meson's of angular momentum $J=j\pm{1\over2}$.  The lowest mass degenerate doublets are
\begin{equation}
\begin{array}{lcl}
  \begin{array}{lcc}
  \left[{1\over2}^-\right]   &   J^P  & (0^-,1^-) 
  \end{array}
&  D,D^* & \mbox{``elastic"} \,, \\[4mm]
  \begin{array}{lcc}
  \left[{1\over2}^+\right]   &   J^P  & (0^+,1^+) \\[2mm]
  \left[{3\over2}^+\right]   &   J^P  & (1^+,2^+)
  \end{array} 
& \Biggr\} & P\mbox{-wave doublets} \,.
\end{array}
\label{degenerate}
\end{equation}
In the HQS limit the semi-leptonic decay $B\to D^{**}\ell\nu$ is determined up to an unknown IW function for each degenerate doublet.  We adopt the notation that $D^{**}$ represents any $D$ state including $D$ and $D^*$.  The decay rate into a single $D^{**}$ is
\begin{equation}
{d\Gamma\over d\omega} (B\to D^{**} \ell\nu) = {G_F^2 |V_{cb}|^2\over 48\pi}
m_B^2 m_{D^{**}}^3 \sqrt{\omega^2-1} f^{**} |\xi^{**}(\omega)|^2 \,,
\label{singleD}
\end{equation}
where the invariant velocity transfer is $\omega=v'\cdot v$ and the $f^{**}$ functions have a known dependence on $w$ and the meson masses\cite{ov1}.  As mentioned previously, the IW function $\xi^{**}(\omega)$ is the same within a degenerate doublet.

Since the IW functions are intrinsically non-perturbative it will be useful to define them in terms of the LDF wavefunctions\cite{zalewski}.  This can be done starting from the defining matrix element which can be enumerated and evaluated using the covariant trace formalism\cite{georgi}
\begin{equation}
   \left< \Psi_{**}(v') |J_\Gamma| \Psi_B(v) \right> =
 {\rm Tr} \left[ \bar M_{**}(v') \Gamma M_B(v) \right] \xi^{**}(\omega) \,.
\label{trace}
\end{equation}
On the other hand HQS requires the meson state to factor into products of the
LDF wavefunction and the heavy quark free bispinor (schematically) as
\begin{equation}
\Psi(v) = \sum_{\lambda\lambda_{Q}} \left<\mbox{Clebsch}\right>
\Phi_{j\lambda_j}(v) u_{\lambda_Q}(v) \,. \label{clebsh}
\end{equation}
Substitution into the left side of (\ref{trace}) then yields a unique IW function in terms of the LDF wavefunction for any transition\cite{ov1}.

\section{ Elastic Decay $[0^-\to 0^-\mbox{ or }1^-]$ }

Decay to a hadron state in the same doublet as the initial $B$ can be thought of as elastic.  These decays form the bulk of the semi-leptonic process as seen from (\ref{Br}).  The IW function in this case is given by
\begin{equation}
 \xi(\omega) = \sqrt{ 2\over\omega+1} \left< \Phi(v') \mid \Phi(v) \right> \,.
\end{equation}
Explicit evaluation of the time and angular integrations is most easily done in the Breit frame\cite{zalewski} as
\begin{equation}
  \xi(\omega) = {2\over\omega+1} \left< j_0 (ar) \right>_{00} \,,
\label{breit}
\end{equation}
where for the radial ground state the parameter $a$ is defined as
\begin{equation}
  a =   2E\sqrt{\omega-1\over\omega+1} \,.
\end{equation}
Since the IW function is model dependent one approach is to expand this function about the hadron zero-recoil point $\omega=1$, leading to the expression
\begin{equation}
 \xi(\omega)= \xi(1)-\rho^2(\omega-1)+c(\omega-1)^2+\cdots \;.
\label{expand}
\end{equation}
Because of axial current renormalization and $1/m$ corrections the constant term is expected to be about 10\% less than the nominal value $\xi(1)=1$.  To approximate
a normal formfactor a curvature term $c$ must be included\cite{burdman} but this extra parameter seriously decreases the statistical accuracy.  A useful step forward has been to try to constrain the curvature parameter in terms of the slope parameter $\rho^2$ using principles of some generality.  One such effort by Caprini and Neubert\cite{cn} used some ideas of HQS together with general properties of analyticity and unitarity to establish the dark shaded constraint region shown in Fig.~1.

\begin{figure}[t]
\centering
\hspace{0in}\epsfxsize=3.5in\epsffile{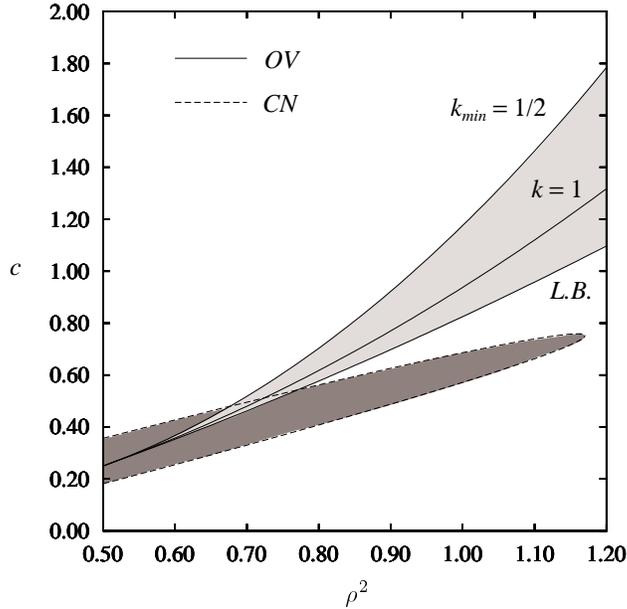}

\medskip
\caption{The allowed range of curvature $c$ for a given Isgur-Wise slope
         parameter $\rho^2$.  The darker shaded region is the result of                                                            
         Caprini and Neubert\protect\cite{cn} and the lighter region is that of Olsson                 and Veseli\protect\cite{ov2}.}
\end{figure}

There are two general comments which one might make about the Caprini-Neubert
approach.  The expansion about the zero recoil point (\ref{expand}) is not strongly convergent in the region where most of the data is found, i.e., $\omega$ near the upper
kinematic limit of about $\omega=1.5$.  Higher coefficients in this expansion would appear difficult to compute.  Another possible problem is the phenomenon of the ``would be anomalous threshold singularity"\cite{jaffe}.  The short range interaction
will give rise to an anomalous singularity below the meson annihilation threshold if confinement is not present.  In an actual confined meson the effects of this singularity may remain even though the absorptive part now
resides at or above the annihilation threshold in a form not easily related to
physical processes.  The ``would be" effect is clearly related to treating the meson as an elementary particle.

Both of the above problems can be avoided by evaluating (\ref{expand}) with the help of (\ref{breit}) in terms of the slope parameter.  From differentiation of (\ref{breit}) we have\cite{ov2}
\begin{eqnarray}
  \rho^2 &= & {1\over2} + {E^2\over3} \left< r^2 \right>_{00} \,,\\
      c  &= & {1\over4} + {E^2\over 3} \left< r^2 \right>_{00} 
+ {E^4\over 30} \left< r^4 \right>_{00}   \,,                           \end{eqnarray}
where $\left<r^2\right>$  and  $\left<r^4\right>$  are expectation values in the ground state.  Eliminating the LDF energy yields our result
\begin{equation}
  c=\rho^2-{1\over4} + {3\over10} \left( \rho^2 - {1\over2} \right)^2 \left( \left< r^4\right> \over \left< r^2 \right>^2 \right) \,.
\end{equation}
Other than the slope parameter the curvature depends only on the ratio
$\left<r^4\right>/\left<r^2\right>^2$.  This ratio has a lower bound of unity from the Schwartz
inequality.  An upper bound is provided from a heavy-light lattice simulation
\cite{duncan} which indicates that the wavefunction is substantially given by a linear
exponential.  We adopt a conservative upper bound corresponding to a wavefunction falling off as $\exp[-\sqrt r \;]$.  The result is shown in Fig.~1 by the lightly shaded region.  We observe that although the two curvatures agree for smaller $\rho^2$ they differ substantially for $\rho^2>1$.

We note, as a concluding comment on elastic decay, that once the heavy-light
wavefunction is known (\ref{breit}) yields the IW function and the D* or D distributions
are then computed using (\ref{singleD}).  Relativistic models, which account for the observed heavy-light states, then make predictions for the experimental distributions which are in good agreement\cite{ov3}.

\section{Inelastic Semi-leptonic Decay}

Inelastic semi-leptonic decays into a single excited $D$ state are also computed using (\ref{singleD}) but with an appropriately defined IW function.  For example the four $P$-wave states mentioned in (\ref{degenerate}) are described by two IW functions:\cite{ov1}
\begin{eqnarray}
  \xi_{{1\over2}^+} (\omega) &=&  {2\over\sqrt{\omega^2-1}} \left< j_1 (ar) \right>_{10} \,,\\
\xi_{{3\over2}^+}(\omega) &=& {4\sqrt3\over(\omega+1)\sqrt{\omega^2-1}} \left< j_1(ar)\right>_{10}
\,, \label{3/2+} 
\end{eqnarray}
where the expectation value is between the initial and final states. From the two LDF energies  $E$ and $E'$ we have  $a=(E+E')\sqrt{\omega-1\over\omega+1}$.
If the spin-orbit interaction is small the $\left[{1\over2}^+\right]$ and $\left[{3\over2}^+\right]$ doublets are degenerate and
\begin{equation}
\xi_{{3\over2}^+} (\omega) = {2\sqrt 3\over\omega+1} \xi_{{1\over2}^+}(\omega) \,.
\label{xi-degen}
\end{equation}

Again using a variety of relativistic models\cite{ov4}, which accurately account for the observed heavy-light spectroscopy, the $P$-wave IW function $\xi_{{3\over2}^+}$ is shown in Fig.~2.

\begin{figure}[t]
\centering
\hspace{0in}\epsfxsize=3.5in\epsffile{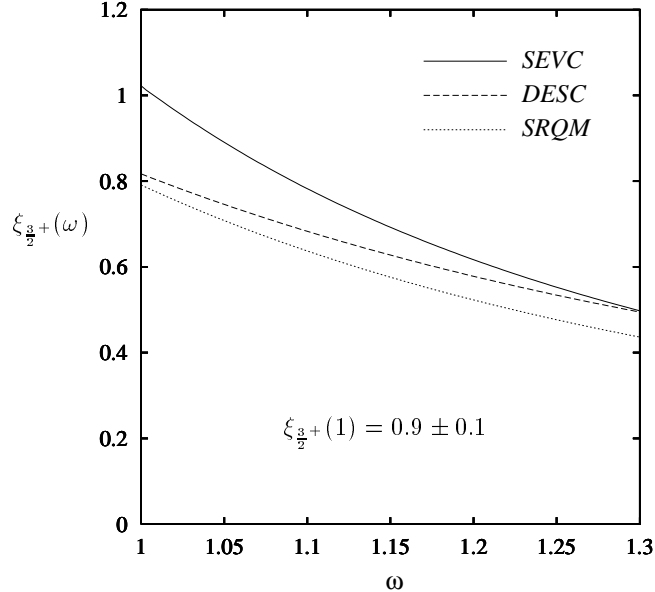}

\medskip
\caption{The Isgur-Wise function $\xi_{{3\over2}^+}$ corresponding to the LDF $j=3/2$. Here\protect\cite{ov4} are shown predictions for the realistic models SEVC (Salpeter Equation Vector Confinement), DESC (Dirac Equation Scalar Confinement), and SRQM (Semi-Relativistic Quark Model) using (\protect\ref{3/2+}).}
\end{figure}

Our predictions as well as two others are given in Table 1.
For each $P$-wave state the numbers represent the expected percent branching ratios. Our results\cite{ov4} assume heavy quark symmetry and the other two contain finite heavy quark masses.  The updated Isgur, Scora, Grinstein, Wise (ISGW2) model is well known\cite{scora} while the more recent work  Leibovich, Ligeti, Stewart, and Wise\cite{leibo} is done in the HQET formalism.

\begin{table}[t]

\caption{Branching ratios in percent for semi-leptonic decay into a $P$-wave excited $D$ state.  The theoretical results are due to                   V.O.\protect\cite{ov4}, ISGW2\protect\cite{scora} and LLSW\protect\cite{leibo}.}
\medskip
\centering
\hspace{0in}\epsfxsize=3in\epsffile{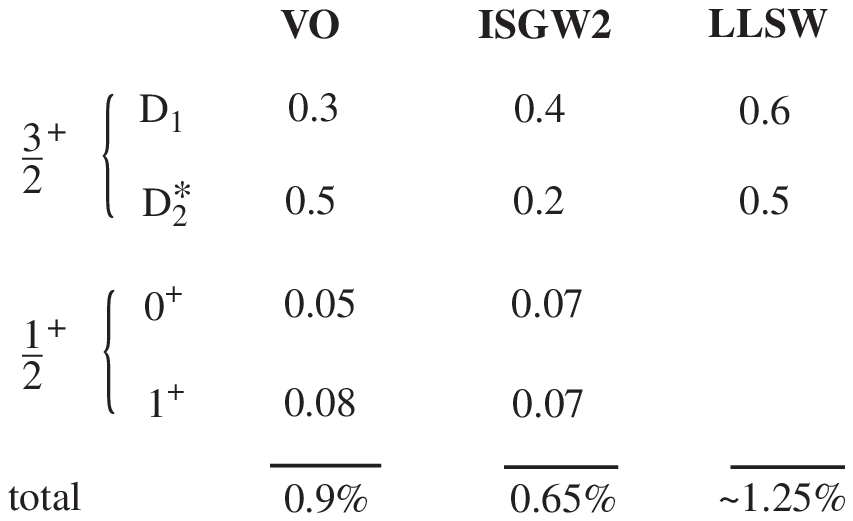}

\end{table}

Some observations on the theoretical results are:
\begin{enumerate}
           \item The results are roughly consistent.
           \item The effect of $1/m$ corrections is to enhance the $D_1$ decay
              at the expense of the $D_2^*$.
           \item The total $P$-wave decay branching fraction is about 1\%.
           \item Semi-leptonic decay into states other than the $D,\,D^*$, or the four $P$-wave states amounts to only 0.1\%.  These include radial
              excitations as well as higher angular momentum states\cite{ov4}.
\end{enumerate}

\section{The Experimental Situation}

There is some experimental evidence concerning the semi-leptonic decay into $P$-wave states.  With the expected Br$(D_1\to D^*\pi)=2/3$ we have,
\begin{equation}
{\rm Br}(B\to D_1\ell \nu)= \left\{
\begin{array}{l}
      (0.49\pm0.13\pm0.06)\%\rm\  CLEO\mbox{\cite{cleo}}\\
      (0.66\pm0.15\pm0.11)\%\rm\  ALEPH\mbox{\cite{aleph}}
\end{array}
\right.
\end{equation}
By comparing to Table 1 one observes that the HQS prediction\cite{ov4} is marginally consistent with the data and the addition of $1/m$ corrections seems to improve
the agreement.  It is important to improve the result for decay into $D_2^*$, which
experimentally has an upper bound of 1\%, to see if the HQET prediction of $D_2^*$
suppression is correct.  This may be best achieved by observing the dominant
$D\pi$ decay of the $D_2^*$.  The ${1\over2}^+$ states will be far more difficult to observe because of their expected large widths and small branching ratios.

We may conclude from the agreement with theory (where available) that HQS/HQET is substantially correct.  Since we have only accounted for about 1\% of the missing $(3.8\pm1.3)\%$ branching ratio (\ref{notsingleD}) we must look elsewhere for the remainder.  Fortunately we have both a valuable experimental clue and a reasonable physical explanation.  The ALEPH group has made the inclusive measurement,
\begin{equation}
{\rm Br}(B\to D\pi\ell\nu+D^*\pi\ell\nu)=(2.3\pm0.3\pm0.3)\%\rm\  ALEPH\mbox{\cite{aleph2}}\,.
\label{aleph}
\end{equation}
This includes decays into $D_1,\, D_2^*$, and the ${1\over2}^+$ doublet states.  If we subtract (\ref{aleph}) from (\ref{notsingleD}) one is left with only $(1.5\pm1.3)\%$ missing.  It seems that the probable explanation of the $B$ semi-leptonic decay channels is that additional pions are emitted when the $B$ decays into the $D$ or $D^*$ states.  This most likely
occurs by string breaking after the decay.  These non-resonant $\pi D$ and $\pi D^*$ states together with the $P$-wave $D$ states account for most of the ``missing" branching ratio.

\section{Sum Rules}

Important consistency relations are provided by the completeness sum rules.  The best known of these is the Bjorken sum rule\cite{bjorken},
\begin{equation}
   \rho^2= {1\over4} + {1\over4} \sum_{{1\over2}^+} \left| \xi_{{1\over2}^+}(1) \right|^2 + {2\over3} \sum_{{3\over2}^+} \left| \xi_{{3\over2}^+}(1)\right|^2 +\cdots 
\;. \label{bjorken}
\end{equation}
The sums are over radial excitations and the ellipsis indicates possible non-resonant hadron decay states.  This sum rule relates the slope of the elastic form factor to the $P$-wave IW functions evaluated at the zero recoil point.  A second well known sum due to Voloshin\cite{voloshin} is,
\begin{equation}
{1\over 2} = {1\over4} \sum_{{1\over2}^+} \left( {E_P\over E_S} - 1 \right) \left| \xi_{{1\over2}^+}(1) \right|^2 + {2\over3} \sum_{{3\over2}^+} \left( {E_P\over E_S} - 1\right) \left| \xi_{{3\over2}^+}(1)\right|^2 + \cdots \;,
\label{voloshin}
\end{equation}
where $E_P$ are radial LDF states and $E_S$ is the $S$-wave ground state.  These two sum rules form a powerful constraint between the form factors and the meson spectroscopy ultimately
relating to the allowed interaction\cite{ov5}.  To illustrate the nature of this constraint let us consider a simplified case which is not too far from reality.  We will assume that:
\begin{enumerate}
          \item  The spin orbit interaction can be neglected compared to the
                  angular momentum excitation.  This implies that the ${1\over2}^+$ and ${3\over2}^+$ doublets are degenerate and (\ref{xi-degen}) is valid.

          \item  Only the ground state radial excitation contributes                  significantly to the sums in (\ref{bjorken}) and (\ref{voloshin}).  This is                  reasonable since the higher radial states will have nodes
                  which cause cancellations in the overlaps.
\end{enumerate}
The two sum rules then become,
\begin{eqnarray}
       \rho^2 &=& {1\over4} + \left| \xi_{{3\over2}^+}(1) \right|^2 \,, \label{sumrules-a}\\
       {1\over2} &=& \left( {E_P\over E_S} - 1\right) \left| \xi_{{3\over2}^+} (1) \right|^2 \,.
\label{sumrules-b}
\end{eqnarray}
If we eliminate the IW function we find
\begin{equation}
  \rho^2 = {E^2_P - E^2_S \over 4 (E_P-E_S)^2} \,.
\label{rho2}
\end{equation}
We assume normal linear Regge trajectories for the LDF.  As shown recently\cite{olsson}, the heavy-light Regge slope is exactly double the light-light slope under fairly general conditions.  This gives
\begin{equation}
   \alpha'_{\rm HL} = {\Delta j\over\Delta E^2} = {1\over E^2_P-E^2_S} =  2\alpha'_{\rm LL} \,.
\label{alpha'}
\end{equation}
Combining (\ref{rho2}) and (\ref{alpha'}) yields our result,
\begin{equation}
\rho^2= \left[ 8\alpha'_{\rm LL} \left(E_P-E_S\right)^2 \right]^{-1}  \,.
\end{equation}
With the well known experimental values $\alpha'_{\rm LL}=0.9\rm~GeV^{-2}$ and $E_P-E_S=0.4$~GeV we find that
\begin{equation}
\begin{array}{rcl}
 \rho^2 &=& 0.9 \,,\\
 \xi_{{3\over2}^+}(1) &=& 0.8 \,,\\
 E_S &=& 0.5\rm~GeV \,,\\
 m_c &=& 1.5\rm~GeV \,.
\end{array}
\end{equation}
All of the above numbers turn out to be reasonable.  In particular we see from Fig.~2 that the inelastic IW function $\xi_{{3\over2}^+}(1)$ at the zero recoil point is consistent with the result of detailed model expectations.

\section{Conclusions}

In this talk I have mostly concentrated on the heavy quark symmetry limit of $B$ semi-leptonic decay.  My first aim was to outline how the formfactors are defined and to indicate that HQS is in good shape so far with $1/m$ corrections
having the right sign to improve agreement with experiment.  In  particular
it will be  important to measure the $D^*_2$ decay to verify that $1/m$ corrections depress this channel as well as enhance the $D_1$ mode.

Another question addressed here is the ``missing" semi-leptonic branching fraction.  Subtracting the $D$ and $D^*$ decays from the inclusive branching fraction leaves over 3\% not accounted for.  The four $P$-wave $D$ mesons contribute about 1\% and all remaining higher radial and angular states add up to only  about 0.1\%.  A recent ALEPH measurement indicates that $D\pi$ and $D^*\pi$ non-resonant decays probably  provide the bulk of the missing branching fraction.

The power of the completeness sum rules of Bjorken and Voloshin was also discussed.  A simple example illustrates how the heavy-light spectroscopy is intimately related to the elastic and inelastic form factors at the hadron zero recoil point.

\section*{Acknowledgments}

I would like to thank Sini\v sa Veseli for helpful conversations. This research was supported in part by the U.S.~Department of Energy under Grant No.~DE-FG02-95ER40896 and in part by the University of Wisconsin Research Committee with funds granted by the Wisconsin Alumni Research Foundation.

\end{document}